\documentclass[11pt]{article}
\usepackage{cite}
\usepackage{amsmath,amsfonts,amssymb}
\usepackage[small,bf,hang]{caption}
\usepackage{slashed}
\usepackage[latin5]{inputenc}

\def\hybrid{
        \topmargin -20pt
        \oddsidemargin 0pt
        \headheight 0pt \headsep 0pt
        \textwidth 6.25in 
        \textheight 9.5in 
        \marginparwidth .875in
        \parskip 5pt plus 1pt \jot = 1.5ex}

\hybrid

 \csname
@addtoreset\endcsname{equation}{section}

\def\moth{\mathsurround=0pt}
\newdimen\zo \zo=0pt

\def\tick{\leaders\hrule height 0.5ex depth 0pt \hskip 0.5pt}
\def\upboxfill{$\moth \setbox\zo\hbox{\tick}%
  \hskip 3pt\hbox to 0pt{$\tick$\hss}\hrulefill \hbox to 7.5pt{$\tick$\hss}$}

\def\dtick{\leaders\hrule height .34pt depth 0.5ex \hskip 0.5pt}
\def\downboxfill{$\moth \setbox\zo\hbox{\dtick}%
  \hskip 2pt\hbox to 0pt{$\dtick$\hss}\hrulefill \hbox to 2pt{$\dtick$\hss}$}

\def\ciftS{\mathbb{S}}
\def\cL{{\cal L}}
\def\cH{{\cal H}}
\def\cK{{\cal K}}

\def\del{\partial}

\def\be{\begin{equation}}
\def\ee{\end{equation}}
\def\bea{\begin{eqnarray}}
\def\eea{\end{eqnarray}}
\def\ba{\begin{array}}
\def\ea{\end{array}}

\begin{document}

\begin{titlepage}
\rightline{} \rightline\today
\begin{center}
\vskip 2.0cm {\Large \bf {Yang-Baxter Deformation as an  O(d,d) Transformation}}\\
\vskip 1.2cm {\large {Aybike \c{C}atal-\"{O}zer}$^{1,\sharp}$ and
Se\c{c}il Tunal\i$^{1,2,\flat}$} \vskip 1cm

{\it $^{1}$ {Department of Mathematics,}}\\
{\it {\.{I}stanbul Technical University,}}\\
{\it {Maslak 34469,
\.{I}stanbul, Turkey}}\\
 \vskip 0.5cm
{\it $^{2}$ {Department of Mathematics,}}\\
{\it {\.{I}stanbul Bilgi University,}}\\
{\it {Beyo\u{g}lu 34440,
\.{I}stanbul, Turkey}}\\
  \vskip 1cm {\bf Abstract}
\end{center}

\vskip 0.1cm

\noindent
\begin{narrower}

\noindent The  rules for Yang-Baxter (YB) deformation for a
generic Green-Schwarz string sigma model has been obtained
recently. We show that the deformation can be described through
the action of a coordinate dependent $O(d,d)$ matrix on the target
space fields both in the NS-NS and the RR sectors, generalizing
previous results. This enables us to show that the YB deformed
fields can be regarded as duality twisted fields in the context of
Gauged Double Field Theory (GDFT). We compute the fluxes
associated with the twist and show that the conditions on the
R-matrix determining the YB deformation give rise to conditions
for the fluxes on the GDFT side. More precisely, we show that YB
deformation is a process which takes a solution of DFT with
geometric flux associated with the isometry group $G$ and deforms
it to another solution of DFT with vanishing R-flux and
non-vanishing Q-flux given by the structure constants of the dual
Lie algebra
 determined by the R-matrix.
We also show that the  non-unimodularity of the R-matrix forces
the generalized dilaton field to pick up a linear dependence on
the winding type coordinates of DFT, implying that the
corresponding target space fields  satisfy the field equations of
DFT in the generalized supergravity frame. This provides a new
perspective on the relation between the non-unimodularity of the
R-matrix, the trace of the Q-flux and the generalized supergravity
equations.

\end{narrower}

\vskip 1.2cm

$^{\sharp}$ozerayb@itu.edu.tr \ \ $^{\flat}$tunali16@itu.edu.tr

\end{titlepage}

\newpage

\tableofcontents

\newpage

\section{Introduction}\label{introduction}

Constructing integrable deformations of string backgrounds
relevant for AdS/CFT correspondence is an active line of research.
An important milestone in this direction was the work of Klimcik
\cite{Klimcik:2002zj}, where he introduced a particular type of
deformation for  Principal Chiral Models (PCM) with simple compact
Lie group as the target manifold. The resulting deformed sigma
model was dubbed Yang-Baxter sigma model, as the deformation is
based on solutions of modified classical Yang-Baxter equation
(mCYBE). The integrability of the YB sigma model was proved later
in \cite{Klimcik:2008eq}. The applicability of YB deformations was
extended to symmetric coset spaces  in \cite{Delduc:2013fga},
which in turn was applied to $AdS_5 \times S^5$ in
\cite{Delduc:2013qra}. The NS-NS sector of the corresponding
deformed supergravity background was found
 in \cite{Arutyunov:2013ega}. Similar deformations for the NS-NS
sector of $AdS_n \times S^n$ supercosets was studied in
\cite{Hoare:2014pna}. The RR sectors of these deformed backgrounds
was worked out in \cite{Lunin:2014tsa}, and was successfully
established for the $n=2$ and $n=3$ cases. The most interesting
$n=5$ case could not be understood fully in that work and the full
Lagrangian for the deformed $AdS_5 \times S^5$ was found later in
\cite{Arutyunov:2015qva}. We should note that such deformations of
the supergravity backgrounds are usually called
$\eta$-deformations.

It is also possible to consider similar deformations based on the
classical Yang-Baxter equation (CYBE), as opposed to mCYBE.
Following the literature, we call the resulting models homogeneous
Yang-Baxter models. Such deformations of the $AdS_5 \times S^5$
string was first studied  in \cite{Kawaguchi:2014qwa}. These
deformations are particularly interesting, as they provide a
natural generalization of the Lunin-Maldacena (LM) deformations
\cite{Lunin:2005jy}, obtained by T-duality-shift-T-duality (TsT)
transformations \cite{Frolov:2005dj}. Indeed, when the R-matrix
(which is an endomorphism on the Lie algebra of the symmetry
group) that solves the CYBE is Abelian, the corresponding
deformation is the LM deformation, as was first suggested in
\cite{vanTongeren:2015soa} and then proved
 in  \cite{Osten:2016dvf}. The relation between TsT
transformations and homogeneous YB deformations was also studied
in the papers \cite{Matsumoto:2014nra}-\cite{Hoare:2016wca}. For
future reference, we note here that the LM deformations can be
generated by the action of a constant $O(d,d)$ matrix on the
initial background with commuting isometries, as was first
observed in \cite{CatalOzer:2005mr}. We will henceforth call this
matrix the TsT matrix.

Given the relation between homogeneous  YB deformation based on an
Abelian R-matrix and the TsT transformation, it is natural to
wonder whether a similar relation to T-duality holds for more
general R-matrices. Indeed, it was conjectured in
\cite{Hoare:2016wsk} that the homogeneous Yang-Baxter model could
 be obtained by applying Non-Abelian T-duality (NATD) to the
original background, with respect to an isometry group determined
by the R-matrix. This conjecture was proved by Borsato and Wulff
in \cite{Borsato:2016pas},\cite{Borsato:2017qsx} for the case of
Principal Chiral Models (PCM)\footnote{See also
\cite{Lust:2018jsx}, where they arrive at the same result by
utilizing the $O(d,d)$ structure of NATD.}. Recently, Borsato and
Wulff extended their work to homogeneous YB deformations of more
general sigma models than PCMs. In their paper
\cite{Borsato:2018idb}, they derived the rules for NATD for a
generic Green-Schwarz (GS) string with isometry group $G$, where
NATD is applied with respect to a subgroup of $G$. Then, by using
the connection between NATD and homogeneous YB models that they
had established in \cite{Borsato:2016pas}, \cite{Borsato:2017qsx},
they proposed the form of homogeneous YB deformation for a generic
GS sigma model and showed that this gave a natural generalization
of the YB deformation of PCMs and supercosets.

Recently in \cite{Catal-Ozer:2019hxw}, the first author showed
that the NATD transformation rules  obtained in
\cite{Borsato:2018idb} for a generic GS string with isometry group
$G$ can be described through the action of a coordinate dependent
$O(d,d)$ matrix. The dependence of the NATD matrix on the
coordinates is determined by the structure constants of the Lie
algebra associated with the isometry group. Viewing NATD this way
made it possible to consider it as a transformation in Double
Field Theory (DFT), a formalism which provides an $O(d,d)$
covariant formulation for effective string actions
\cite{Hull:2009mi}-\cite{Hohm:2011dv}. This made it  possible to
prove that the NATD fields (including those in the RR sector)
solve supergravity equations when the isometry group is
unimodular, whereas in the non-unimodular case they form a
solution for generalized supergravity equations (GSE)
\cite{gse1},\cite{gse2}. NATD (and the related Poisson-Lie
T-duality) has been studied in the context of DFT also in the
papers \cite{Lust:2018jsx},\cite{Hassler:2017yza},
\cite{Demulder:2018lmj},\cite{Sakatani:2019jgu},\cite{Thompson:2019ipl}
and \cite{hassler}.

Given the relation between NATD and homogeneous YB deformations
discussed briefly above, the results of \cite{Catal-Ozer:2019hxw}
implies that it should be possible to describe homogeneous YB
deformation also as an $O(d,d)$ transformation.  The purpose of
this paper is to show that this is indeed the case, for the YB
deformation of \emph{any} Green-Schwarz sigma model. More
precisely, we will examine the homogeneous YB deformation proposed
in \cite{Borsato:2018idb} for a generic GS sigma model and show
that the deformation in the target space, including the RR fields
is generated by a particular \emph{non-constant} $O(d,d)$
transformation, called the $\beta$-transformation. We call the
matrix associated with this transformation the Yang-Baxter (YB)
matrix. The R-matrix that determines the YB matrix is constructed
by picking up a fixed, \emph{constant} R-matrix satisfying the
CYBE and extending it by the adjoint action to the whole group
manifold associated with the isometries of the background. The YB
matrix we find in this paper is related to the NATD matrix we
studied in \cite{Catal-Ozer:2019hxw} by a constant $O(d,d)$
transformation. That the two matrices are related is of course not
surprising at all, given the fact that \cite{Borsato:2018idb}
constructed the YB deformed models  by utilizing the connection
with NATD. We will discuss this relationship in section
\ref{ybmatrixsection}. Then, in section \ref{YBLM} we will explore
the relationship between the YB matrix and  the TsT matrix.  We
will show that, in the special case when the R matrix is Abelian,
the YB matrix becomes constant and reduces to the TsT matrix. This
directly follows from the fact that the adjoint action is trivial
on an Abelian Lie algebra, as we will discuss in subsection
\ref{YBLM}.

Viewing YB deformation as an $O(d,d)$ transformation in the target
space is quite useful. First of all, it makes the calculations
considerably simpler. Moreover, the transformation rules for the
fields in the RR sector almost come for free. Computing the RR
fields of the (homogeneous) YB models by the supercoset
construction as in \cite{Arutyunov:2015qva},\cite{Kyono:2016jqy}
is usually cumbersome. An alternative way is to start with the
deformed NS-NS fields and determine the deformed RR fields by
directly solving the Type IIB supergravity equations as in
\cite{Lunin:2014tsa}.\footnote{In the papers
\cite{Araujo:2017jap},\cite{Araujo:2017enj}, the open-closed
string map associated with the YB deformation is extended to the
RR sector by utilizing the Page forms.} Viewing YB deformation as
a $\beta$-transformation  gives a more direct and easy to apply
formula to determine the RR fields in the deformed background,
since the $O(d,d)$ transformation of the NS-NS fields dictates the
transformation for the RR fields. As we discuss in section
\ref{RRfields}, the deformation in the RR sector is completely
determined by the $Spin(d,d)$ matrix $S_{{\rm YB}}$ projecting to
the YB matrix $T_{{\rm YB}}$ as $\rho(S_{{\rm YB}}) = T_{{\rm
YB}}$, where $\rho$ is the usual double covering homomorphism
between $Pin(d,d)$ and $O(d,d)$. We will discuss this in detail in
section \ref{RRfields}. Another advantage of viewing YB
deformation as an $O(d,d)$ transformation is that it gives a
natural embedding of the homogeneous YB model in DFT. As we will
discuss in section \ref{YBinGDFT}, this provides a direct
correspondence between the properties of the R-matrix and the
deformations of supergravity. We will see that the deformations in
the gravity picture, when embedded in DFT, will be encoded in
entities called fluxes, which deforms DFT to Gauged Double Field
Theory (GDFT),
\cite{Geissbuhler:2011mx},\cite{Aldazabal:2011nj},\cite{Grana:2012rr},\cite{Catal-Ozer:2017cqr}.
Then, the conditions to be satisfied by the R-matrix  manifest
themselves as conditions on  these fluxes. More precisely, we will
see that the  R-matrix satisfies the CYBE if and only if the
R-flux vanishes on the DFT side. On the other hand, the structure
constants of the dual Lie algebra determined by the R-matrix
appears as components of the so called Q-flux on the DFT side, and
the R-matrix is unimodular if and only if the Q-flux is traceless.
When the R-matrix is non-unimodular, it is known that the YB
deformed target space fields  constitute a solution to the
equations of generalized supergravity \cite{Borsato:2018idb}. In
section \ref{dilaton}, we look at how Q-flux with non-vanishing
trace deforms DFT. We see that the resulting GDFT is consistent
only when the generalized dilaton field picks up a linear
dependence on the winding type coordinates in a particular way,
which does not violate  the section condition of DFT. It is known
that the field equations of DFT reduce to generalized supergravity
equation for this particular solution of the section constraint,
\cite{Sakatani:2016fvh}, \cite{Baguet:2016prz},
\cite{Sakamoto:2017wor}. Therefore, our approach provides a new
perspective on the  connection between the non-unimodularity of
the R-matrix, the trace of the Q-flux and generalized supergravity
equations.

The $O(d,d)$ structure of YB deformations has already been studied
by several groups. Let us review briefly what has been achieved in
those works and discuss what has been added in our paper.  The
first paper, where it was proposed that the  homogeneous YB
deformation could be obtained by the action of a $\beta$-
transformation is \cite{Sakamoto:2017cpu}. This proposal is based
on observation and is supported by working out the example  of the
$AdS_5 \times S^5$ background. Later in \cite{Sakamoto:2018krs}, a
proof of the proposal of \cite{Sakamoto:2017cpu} was given for
deformations of the $AdS_5 \times S^5$ background, by rewriting
the YB deformed action in the form of Green-Schwarz action (up to
quadratic order in fermions) and showing that the target space can
be obtained by applying $\beta$-transformation on the $AdS_5
\times S^5$ background. It should be noted that their formulas
worked only when the B-field of the undeformed background
vanished. Indeed,  they also considered $\beta$-transformations of
the $AdS_3 \times S^3 \times T^4$  with non-vanishing H-flux and
showed that the resulting background solved equations of motion of
(generalized) supergravity. However, a proof for the equivalence
with YB deformation could not be given for this case, due to
complications arising from the presence of the H-flux. Our work
extends the work of \cite{Sakamoto:2018krs} in two ways. Firstly
and most importantly, the equivalence of YB deformations and
$\beta$ transformations is not shown merely on a case by case
basis and works for the generic case (of which $AdS_5 \times S^5$
is an example). This is because, the YB deformation rules
presented in \cite{Borsato:2018idb} (and which we rewrite here as
an $O(d,d)$ transformation) is generic and works for any GS sigma
model. Secondly, in our case, we do not need to take the B-field
to vanish in the initial background. It is also important to note
that our approach in  this paper is different from the approach of
\cite{Sakamoto:2017cpu},\cite{Sakamoto:2018krs} in that, here we
use the framework of GDFT and interpret the $\beta$-transformation
associated with the YB matrix as a solution generating
transformation, which takes one solution of DFT with geometric
flux to another one with
 dual Q-flux, as we will explain in detail in section
 \ref{YBinGDFT}. In contrast, the  Q-flux
of  the YB deformed model cannot be captured in the papers
\cite{Sakamoto:2017cpu},\cite{Sakamoto:2018krs}. Later in
\cite{Fernandez-Melgarejo:2017oyu}, it is shown that the deformed
backgrounds found in
\cite{Sakamoto:2017cpu},\cite{Sakamoto:2018krs} belong to a
certain class of non-geometric backgrounds, called T-folds
\cite{Hull:2004in}, and the Q-flux of the YB deformed backgrounds
is understood as arising from a non-trivial monodromy in the so
called $\beta$-field around closed cycles. Our findings in this
paper, which we obtain by
 computing directly the Q-flux associated with
the YB matrix in the framework of GDFT  agree with the results of
\cite{Fernandez-Melgarejo:2017oyu} and  extends it in the sense
that it reveals the role of Q-flux as
 the dual structure constants to geometric flux. Again, it is worth noting that the monodromy of
the $\beta$-field is computed in
\cite{Fernandez-Melgarejo:2017oyu} for the Minkowski and the
$AdS_5 \times S^5$ background and for vanishing B-field, whereas
our results in this paper applies to any GS sigma model with
isometries. The  Q-flux of deformed backgrounds and its role as
 the dual structure constants to geometric flux is
 also studied in \cite{Lust:2018jsx}, but in the context of Poisson Lie
 duality of sigma models. However, for the YB deformation, this
 feature of the deformed models cannot be captured in
 \cite{Lust:2018jsx}, either.

We should also note that the approach adopted in the papers
\cite{Araujo:2017jap},\cite{Araujo:2017enj},\cite{Araujo:2017jkb},\cite{Araujo:2018rho},
which regards YB deformation as an open-closed string map
\cite{Seiberg:1999vs} also contributes to the understanding of the
$O(d,d)$ structure of YB deformations. Indeed, as we will discuss
in subsection \ref{ybmatrixsection}, the open-closed string map
can be regarded as a special case of $\beta$-transformation. In
this picture, the YB deformation is encoded in the
non-commutativity parameter $\Theta$ of the open-closed string
map\footnote{We would like to note that  the papers
\cite{Tongeren1, Tongeren2} also study the relation between YB
deformations and non-commutative deformations within the context
of AdS/CFT duality.}. In the papers
\cite{Bakhmatov:2017joy}-\cite{Araujo:2018rbc}, the open-closed
string map was promoted to a solution generating symmetry in
generalized supergravity and it was shown that the field equations
enforce the CYBE for the antisymmetric bivector $\Theta$.
Following this line of work which regards YB deformation as an
open-closed string map, the relation between deformations of
supergravity backgrounds and solutions of the CYBE was further
explored in \cite{Bakhmatov:2018bvp} in the context of $\beta$
supergravity.

The plan of the paper is as follows. In section \ref{section1}, we
find  the coordinate dependent $O(d,d)$ matrix $T_{{\rm YB}}$,
which we call the YB matrix, that encodes the YB deformation
obtained in \cite{Borsato:2018idb} of  a generic GS sigma model.
Also in the same section, we discuss how this matrix reduces to
the $O(d,d)$ matrix associated with Lunin-Maldacena deformations,
when the R-matrix  is Abelian. In section \ref{RRfields}, we
describe the deformation in the RR sector, by identifying the
$Spin(d,d)$ matrix that projects to $T_{{\rm YB}}$ under the
double covering homomorphism between $O(d,d)$ and $Pin(d,d)$. The
purpose of section \ref{YBinDFT} is to describe the $O(d,d)$
transformation that generates the  YB deformation as a solution
generating transformation in DFT. After giving a quick review of
DFT, we describe in section \ref{YBinGDFT} the YB deformed fields
as twisted fields in GDFT. In section \ref{fluxcomp} we compute
the fluxes associated with the twist matrix $T_{{\rm YB}}$ and
study the relation between the conditions on the R-matrix and
those on fluxes. As a preparation for this computation, we discuss
some properties of the R-matrix and the dressed R-matrix in
\ref{rmatrix}. Then in section \ref{dilaton} we discuss how the
non-unimodularity of the R-matrix forces the generalized dilaton
field to have a linear dependence on winding type coordinates so
that the YB deformed fields form a solution of DFT in the
generalized supergravity frame. We conclude with discussions and
outlook in section \ref{conclusion}. The paper is complemented
with three appendices.

\section{ Yang-Baxter Deformation as an $O(d,d)$
transformation}\label{section1}

As we discussed in section \ref{introduction} above, it was shown
in \cite{Sakamoto:2018krs} that the YB deformation of the string
sigma model with target space  $AdS_5 \times S^5$ is produced by
the action of a coordinate dependent $O(d,d)$ transformation (a
particular type of $\beta$-transformation, see (\ref{YBmatrix}))
on the fields in the target space. Here, we extend their results
and show that the correspondence holds for the homogeneous YB
deformation of any GS string sigma model, of which $AdS_5 \times
S^5$ is a specific example.  This correspondence could not be
checked in full generality previously, mainly due to complications
in formulating the homogeneous YB deformations of type II
superstring on backgrounds with non-vanishing H-flux. This was
achieved  in \cite{Borsato:2018idb}, where they also checked the
equivalence to the $\beta$-transformation  of
\cite{Sakamoto:2018krs} for the background $AdS_3 \times S^3
\times T^4$ with non-vanishing H-flux. Here, we show the
equivalence of the YB deformation rules found in [24] with the
$\beta$-transformation rules for \emph{any} GS string sigma model,
without having to refer to specifics of the initial supergravity
background.

In order to determine the  the $O(d,d)$ matrix that produces the
YB deformation rules found in \cite{Borsato:2018idb}, analyzing
the deformation in the NS-NS sector is sufficient. Once we find it
from this analysis, the transformation of the RR fields is
determined immediately. In Abelian T-duality, RR fields are
packaged in a differential form, which can be a regarded as a
spinor field that transforms under $Spin(d,d)$. If the fields in
the NS-NS sector transform under $T \in O(d,d)$, then the spinor
field that encodes the RR fields transform under $S_T \in
Spin(d,d)$, which is the element that projects onto $T$ under the
double covering homomorphism $\rho$ between $O(d,d)$ and
$Spin(d,d)$, that is, $\rho(S) = T$ \cite{Fukuma}. Similarly, once
we figure out the $O(d,d)$ matrix that produces the YB
deformation, the transformation of the RR fields is automatically
determined by the corresponding $Spin(d,d)$ matrix, as in Abelian
T-duality. We will discuss this in section \ref{RRfields}.

\subsection{The YB matrix}\label{ybmatrixsection}

The  rules for NATD for a generic Green-Schwarz string sigma model
with isometry group $G$ has recently been obtained in
\cite{Borsato:2018idb}. In \cite{CatalOzer:2005mr}, we identified
the  matrix whose action on the target space produces the NATD
background presented in \cite{Borsato:2018idb}. We call this
matrix the NATD matrix. It is an $O(10,10)$ matrix and is obtained
by embedding (as in (\ref{embedding}) in Appendix \ref{app2}) the
following $O(d,d)$ matrix $T$ in $O(10,10)$: \be
\label{NATDmatrix} T_{{\rm NATD}} =
\left(\begin{array}{cc} 0 & 1 \\
                        1 & \theta_{IJ} \end{array}\right). \ee
                        Here, $\theta_{IJ}$ is given as $\theta_{IJ}
                        = \nu_K C_{IJ}^{\ \ K}$, where $C_{IJ}^{\ \
                        K}, I, J, K = 1, \cdots, d = {\rm dim} G$ are the structure constants of the Lie
                        algebra of the isometry group $G$ of the
                        initial background, and $\nu_I$ are the
                        coordinates of the NATD background. It was
                        shown in \cite{CatalOzer:2005mr} that this matrix
                        gives the correct transformation rule for
                        the fields both in the NS-NS and the RR sector.

In the same paper \cite{Borsato:2018idb}, the YB deformation for a
generic GS sigma model was  proposed by exploiting the relation
between NATD and YB deformations
\cite{Hoare:2016wsk},\cite{Borsato:2016pas},\cite{Lust:2018jsx}.
The deformed model they construct is generated by solutions of
classical Yang-Baxter equation and it generalizes the construction
for PCM and supercosets. The formulas they find for the deformed
metric and the B-field are given in equations (4.12)-(4.13) in
their paper. As noted in the  paper, these transformation rules
can be combined in  one compact rule, which we present
below\footnote{Note that due to the convention in
\cite{Borsato:2018idb} there is a difference in the sign in front
of the B field term.}: \be \label{YBrule} G' + B' = (G + B) [1 +
\eta R_g (G +B)]^{-1}.\ee Here $\eta$ is the deformation
parameter, and $R_g$ is the dressed R-operator, related to the
R-matrix $R$ as  \be \label{dressed} R_g = Ad_{g^{-1}} R Ad_g, \ee
that is, \be \label{dressed2} R_g (X) = g^{-1} R(g X g^{-1}) g, \
\ g \in G, \ \ X \in \mathbf{g}, \ee where $\mathbf{g}$ is the Lie
algebra of the isometry group $G$. The relation between the NATD
and YB parameters is given in \cite{Borsato:2018idb} as \be
\label{parameters} \nu_K C_{IJ}^{ \ \ K} = \eta^{-1} R_{IJ}^{-1} -
\eta^{-1} (R_g^{-1})_{IJ}. \ee  Here, the transformation
(\ref{YBrule}) acts on the metric $G(x)$ and the B-field $B(x)$
defined in (\ref{denklem1}), that is, it acts on $G_{IJ}$ and
$B_{IJ}$ with Lie algebra indices. Note that $l^I_{\ i} G_{IJ}
l^J_{\ j} = G_{ij}$ and similarly for $B_{IJ}$, where $l^I_{\ i}$
are the functions on $G$ that determine the left-invariant
one-forms $\sigma^I = l^I_{\ i} d\theta^{i}$. For more details on
the index conventions, see Appendix (\ref{conventions}).

As discussed in \cite{Borsato:2018idb}, the transformation
(\ref{YBrule}) can be seen as a generalization of the open-closed
string map of
\cite{Araujo:2017jkb},\cite{Araujo:2017jap},\cite{Araujo:2017enj}.
The best way to see this is to realize that \be
\label{rightinvunderleft} (Ad_{g^{-1}})_I^{J} = K_I^{J}, \ee where
we have defined \be \label{Kdefn} K_I^{J} = k_I^{\ \mu} l^J_{\
\mu}. \ee Here, $l^I_{\ \mu}$ are as above and $k_I^{\ \mu}$ are
the components of the right invariant vectors fields $k_I$: \be
\label{rightinvvf} k_I = k_I^{\ \mu} \del_{\mu},\ee that satisfies
\be\label{liealg} [k_I, k_J] = -C_{IJ}^{\ \ K} k_{K}. \ee These
are Killing vector fields for the left invariant metric $G_{\alpha
\beta}$ in (\ref{metric3}). The relation (\ref{rightinvunderleft})
arises from the well-known relation \be r_g^{\ *} \sigma =
(Ad_{g^{-1}}) \sigma \ee where the left hand side corresponds to
the pull-back of the Maurer-Cartan form $\sigma = \sigma^I T_I =
l^I_{\ i} d\theta^{i} T_I$ under the right translation $r_g$,
generated by $k_I$. Then the dressed R-operator in (\ref{dressed})
can be written as \be (R_g)^I_{\ L} =  K^I_{J} R^{J}_{K}
\tilde{K}^K_{L}, \ee where $ \tilde{K}^I_{J} = (Ad_g)^I_{J}$.
Raising the indices of the left hand side with the Cartan-Killing
metric $\kappa^{LM}$ we obtain \bea \nonumber (R_g)^{IM} &\equiv &
(R_g)^I_{\ L} \kappa^{LM} = K^I_{J} R^{J}_{\ K} \tilde{K}^K_{L}
\kappa^{LM}
\\ \nonumber
& = & K^I_{J} R^{J}_{\ K} (\tilde{K}^{-1})^M_{L} \kappa^{LK} \\
& = & \label{dressed2} K^I_{J} R^{JL} K^M_{L}. \eea In the last
line above we used $\tilde{K} = K^{-1}$, which follows from
$Ad_{g^{-1}} = Ad_g^{-1}.$ Also, in the second line we used the
fact that the Cartan-Killing metric is $Ad$-invariant, that is,
$Ad_g$ is orthogonal with respect to $\kappa$, which implies that
\be \tilde{K}^K_{L} \kappa^{LM} = (\tilde{K}^{-1})^M_{L}
\kappa^{LK}.\ee Equivalently \be \tilde{K}^K_{L} \kappa^{LM}
\tilde{K}^R_{M} = \kappa^{RK}. \ee Since the matrix $l^I_{\ \mu}$
is invertible, we can use its inverse $l^{\mu}_{\ I}$ to convert
all algebra indices $I$ to curved indices $\mu$. Then, acting on
the curved background $G_{\mu \nu}, B_{\mu \nu}$ in
(\ref{metric1}) the YB transformation rule is again as in
(\ref{YBrule}) with \be \label{killing}  (R_g)^{\mu \nu} = R^{IJ}
k_I^{\ \mu} k_J^{\ \nu}. \ee This is indeed  a generalization of
the open-closed string map in the papers \cite{Bakhmatov:2017joy},
\cite{Bakhmatov:2018apn} as it reduces to it for vanishing
B-field\footnote{It also agrees with the transformation rules
given in \cite{Bakhmatov:2018bvp} for non-vanishing B-field.}.
Indeed for vanishing B-field the transformation rule
(\ref{YBrule}) reduces to \be \label{onemlidipnot} G' + B' = G[1 +
\eta R_g \ G]^{-1} = [G^{-1} + \eta R_g]^{-1},\ee which is the
same as equation (1.1) of \cite{Bakhmatov:2018apn} with
$\Theta_{NC} = \eta R_g$, where $\Theta_{NC}$ is  the
noncommutativity parameter\footnote{Note that the noncommutativity
parameter $\Theta_{NC}$ is identified as $\Theta_{NC} = -2 \eta'
R_g$ in the papers \cite{Bakhmatov:2017joy},
\cite{Bakhmatov:2018apn}. Therefore the deformation parameter
$\eta$ in our paper and $\eta'$ in these papers are related as
$\eta = -2 \eta'$.}. Note that the bi-Killing anzats they
introduce in equation (1.2) of \cite{Bakhmatov:2018apn} is the
same as what is given in (\ref{killing}).

In order to recast the transformation rule (\ref{YBrule}) as an
$O(d,d)$ transformation, we use the terminology discussed in
\cite{Catal-Ozer:2019hxw} and write $E = G + B$, which is known as
the background matrix \cite{Giveon:1994fu}. It is known that the
action of $O(d,d)$ on the background matrix is by fractional
linear transformations: \be E \rightarrow E' =  T . E = (a E +
b)(c E + d)^{-1},\ee  where \be T =
\left(\begin{array}{cc} a & b \\
                        c & d \end{array}\right) \in O(d,d). \ee
                         Then, (\ref{YBrule}) can be written as
\bea \label{transformE} E' & = & E [\eta R_g \ E + 1]^{-1} \nonumber \\
        & = & T_{{\rm YB}} \ . \ E, \eea where \be
\label{YBmatrix} T_{{\rm YB}} =
\left(\begin{array}{cc}1 & 0 \\
                         \Theta & 1 \end{array}\right). \ee
                        Here, for simplicity, we have defined\footnote{\label{onemlidipnot2} Note that this means $\Theta$ is in fact equivalent to
                        $\Theta_{NC}$ of \cite{Bakhmatov:2017joy}, \cite{Bakhmatov:2018apn}, see equation (\ref{onemlidipnot}) and the discussion
                        below it. However, we call it $\Theta $  rather than $\Theta_{NC}$  as the focus of this paper is not
                        on the relation to noncommutative deformations.}  \be
                        \label{teta}
                        \Theta^{IJ} = \eta (R_g)^{IJ}. \ee
                        This matrix acts only on the
                        isometry directions indexed by $I, J$.
                        The $O(10,10)$ matrix which acts on the
                        full metric and the B-field is obtained by
                        embedding the $O(d,d)$ matrix
                        (\ref{YBmatrix}) in $O(10,10)$ as in (\ref{embedding}). When this is done, one gets exactly the deformation rules presented
                        in equations (4.12)-(4.13) in \cite{Borsato:2018idb} . For simplicity,
                        we will call both matrices (both the $O(d,d)$ and the $O(10,10)$ one) the YB matrix,  denoted by $T_{{\rm YB}}$.

It is well known that the transformation of the background matrix
$E \rightarrow E' = T. E$ under $T \in O(D,D)$ is equivalent to
the transformation \be \label{transformH}
 \cH ~~  \longrightarrow \cH^{\prime} \ = \ T\, \cH
 \,T^T\;,
 \ee where \begin{equation}\label{genmetric}
 \cH =   \left(\begin{array}{cc}
                        G - B G^{-1} B & B G^{-1} \\
                       -G^{-1} B   &  G^{-1} \\
                         \end{array}\right).
\end{equation} In the context of Double Field Theory (DFT), $\cH$ is called the generalized metric.
 Taking $T = T_{{\rm YB}}$, one immediately sees that the
 transformation (\ref{transformH}) and hence the transformation
 (\ref{transformE}) (which is equivalent to (\ref{YBrule})) is the
 same as the $\beta$-transformation of \cite{Sakamoto:2017cpu},\cite{Sakamoto:2018krs}.

 The YB matrix (\ref{YBmatrix}) also produces the correct
 transformation rule for the dilaton field. Under $O(d,d)$ the dilaton field $\phi$ transforms as \be \label{dilaton1} e^{-2 \phi'} = e^{-2 \phi}
\sqrt{\frac{{\rm det }G}{{\rm det }G'}},  \ee where the deformed
metric $G'$ is read off
 from the symmetric part of the deformed background matrix $E'$,
 that is \be G' = \frac{E' + E'^{T}}{2}. \ee It is a well known fact that the transformation (\ref{transformE})
implies for $G'$ the following \cite{Giveon:1994fu}: \be G' =
\frac{1}{(c E + d)^T} \ G \ \frac{1}{(c E + d)}. \ee Then, \be
\frac{{\rm det} G}{{\rm det} G'} = \left({\rm det}(c E +
d)\right)^2. \ee  When $T$ is as in (\ref{YBmatrix}) this gives
\be \left(\frac{{\rm det} G}{{\rm det} G'}\right)^{1/2} = {\rm
det} (\eta R_g (G + B) + 1). \ee Plugging this in (\ref{dilaton1})
gives \be \label{dilaton2} e^{-2 \phi'} = e^{-2 \phi}  {\rm det}
(\eta R_g (G + B) + 1), \ee which is equivalent to \be
\label{dilaton3} \phi ' = \phi - \frac{1}{2} \ln {\rm det} (\eta
R_g (G + B) + 1). \ee This is the same result that has been
obtained in \cite{Borsato:2018idb}.

\subsection{YB deformation as a generalization of the
Lunin-Maldacena deformation}\label{YBLM}

Before we move on to the transformation for the RR fields, let us
explain how the YB matrix  reduces to the matrix that generates
the TsT transformation, when the R-matrix is Abelian. This matrix,
which we will
 call the TsT matrix, was  found in \cite{CatalOzer:2005mr} and it generates the $O(10,10)$ transformation associated
 with  single and multi-parameter Lunin-Maldacena deformations of
\cite{Lunin:2005jy},\cite{Frolov:2005dj}. It is obtained by
embedding the following $O(d,d)$ matrix in $O(10,10)$ as in
(\ref{embedding}): \be \label{LMmatrix} T_{{\rm LM}} =
\left(\begin{array}{cc}1 & 0 \\
                        \Gamma^{IJ} & 1 \end{array}\right). \ee
                        Here $\Gamma^{IJ}$ is antisymmetric with
                        $I, J = 1, \cdots, d$, where $d$
                       is the dimension of the Abelian isometry group
                        $U(1)^d$. For single parameter
                        deformations, the components of $\Gamma^{IJ}$ either vanish or  $\Gamma^{IJ}
                        = -\Gamma^{JI} =
                        \lambda$ for non-vanishing components. For multi-parameter
                        deformations, $\Gamma^{IJ}$ can be equated to
                        different deformation parameters. In this
                        way, one can introduce a total of
                        $d(d-1)/2$ independent deformation
                        parameters \cite{CatalOzer:2009xd}.

Now, assume that the R-matrix $R$ is Abelian. This means that the
Lie subalgebra on which $R$ acts as a non-degenerate operator is
Abelian\footnote{It is known that skew-symmetric solutions for
CYBE on a finite dimensional Lie algebra $\mathbf{g}$ are in
one-to-one correspondence with quasi-Frobenius subalgebras of
$\mathbf{g}$. On such a subalgebra the R-matrix acts as a
non-degenerate operator. If this subalgebra is Abelian, the
R-matrix is  called Abelian; if this subalgebra is unimodular, the
R-matrix is  called unimodular. The rank of the R-matrix is equal
to the dimension of the quasi-Frobenius subalgebra. A
quasi-Frobenius algebra (also called symplectic) is a Lie algebra
which admits a non-degenerate 2-cocycle. See
\cite{Borsato:2016ose} and the references therein and also the
Proposition 22.6 and Proposition 3.1.6 in the book
\cite{Chari:1994pz}.}. From the NATD perspective, the subgroup
$\tilde{G}$ of $G$ with respect to which the NATD has been
performed is Abelian. Restricting to this commutative subgroup and
subalgebra, one immediately sees that $R_g = R$ in
(\ref{dressed}). This is because for a commutative group (assuming
that it is also connected), the adjoint representation is the
trivial representation, that is, the adjoint operator $Ad_A$ acts
as the identity operator on the Abelian Lie algebra of
$\tilde{G}$, for all $A \in \tilde{G}$. As a result, one sees that
for the deformation parameter (or the non-commutativity parameter
in the language of \cite{Araujo:2017jkb}, \cite{Araujo:2017enj})
we have $\Theta^{IJ} =\eta  R^{IJ} = $ constant . This can also be
seen from (\ref{parameters}). Since the NATD group $\tilde{G}$ is
Abelian, the left hand side of (\ref{parameters}) will vanish,
giving $R_g = R$. As a result, for Abelian R-matrices, the
antisymmetric deformation parameter $\Theta^{IJ}$ does indeed have
constant components and the matrix (\ref{YBmatrix}) indeed reduces
to the TsT matrix (\ref{LMmatrix}).

\section{Transformation of  the RR fields}\label{RRfields} Realizing the
deformation of the metric and the B-field as an $O(d,d)$
transformation\footnote{To be precise, the transformation is an
$O(10,10)$ transformation. However, it is obtained by embedding an
$O(d,d)$ matrix in $O(10,10)$ as in (\ref{embedding}), where $d$
is the dimension of the isometry group. Hence the only non-trivial
action is on the isometry directions. For this reason, we usually
refer to this action as an $O(d,d)$ transformation, rather than as
an $O(10,10)$ transformation.} dictates the transformation for the
RR fields immediately. In Abelian T-duality, RR fields are
packaged in a differential form, which can be a regarded as a
spinor field that transforms under $Spin(d,d)$. If the fields in
the NS-NS sector transform under $T \in O(d,d)$, then the spinor
field that encodes the RR fields transform under $S_T \in
Spin(d,d)$, which is the element that projects onto $T$ under the
double covering homomorphism $\rho$ between $O(d,d)$ and
$Pin(d,d)$, that is, $\rho(S) = T$ \cite{Fukuma}. This extends to
the case of NATD, that is,  the transformation of the RR fields is
automatically determined by the $Spin(d,d)$ matrix corresponding
to the NATD matrix, which had been determined in
\cite{Catal-Ozer:2019hxw}. So, it is natural to propose that the
RR fields in the YB deformed background should be produced by the
action of the $Spin(d,d)$ matrix $S_{{\rm YB}}$ projecting to the
YB matrix $T_{{\rm YB}}$: $\rho(S_{{\rm YB}}) = T_{{\rm YB}}$. Our
results should agree with the results found in
\cite{Sakamoto:2018krs}, where  the transformation of the RR
fields is determined by utilizing the methods in
\cite{Hassan:1999bv}. Here, we adopt  the formalism of
\cite{Fukuma}, and the two approaches are known to produce the
same results for the case of Abelian T-duality.

Since $T_{{\rm YB}}$ generates a $\beta$-transformation, finding
the $Spin(10,10)$ element $S_{{\rm YB}}$ that projects onto it is
rather easy. It is given as \be \label{spinelembeta} S_{{\rm YB}}:
\ \ \alpha \longmapsto  e^{ \Theta} \alpha = (1 + i_{\Theta} +
\frac{1}{2}  i_{\Theta}^2 + \cdots ) \alpha, \ee where $\alpha$ is
a generic spinor field (regarded as a differential form through
the Clifford map as in (\ref{spinorF}) and (\ref{cliffordF})
below) and $i_{\Theta} \alpha = \frac{1}{2} \Theta^{IJ} i_{\psi_I}
(i_{\psi_J} \alpha),$ with $i$ being the usual contraction with
$i_{\psi_I} \psi^J = \delta_I^{\ J}$. An important feature of
$S_{{\rm YB}}$ we should note here is that it is an element of
$Spin^+(10,10)$, which is the subgroup of $Spin(10,10)$ connected
to the identity component. This is in contrast with the fact that
$S_{{\rm NATD}}$ is an element of $Spin^-(10,10)$.

 If $F$ is the spinor field that encodes the RR
fluxes of the initial background (assuming that it respects the
isometry $G$), the transformed RR fluxes are read off from the
spinor field $F'$: \be \label{transformF} F' = e^{-B'} S_{{\rm
YB}} . e^{B} F, \ee where $B$ is the B-field of the initial
background and $B'$ is read off from the anti-symmetric part of
$E'$ in (\ref{transformE}): \be B' = \frac{E' - E'^T}{2}. \ee For
more details, we refer the reader to
\cite{CatalOzer:2005mr},\cite{Catal-Ozer:2017cqr} and
\cite{dftRR}.

An important difference from the Abelian case is the following.
Let us write the differential form $F$ as \be \label{spinorF} F =
\sum_p G^{(p)} = \sum_p \left(F^{(p)} + F_I^{(p-1)} \sigma^I +
\frac{1}{2} F_{IJ}^{(p-2)} \sigma^I \wedge \sigma^J + \cdots +
F^{(p-d)} \sigma^1 \wedge \cdots \wedge \sigma^d \right), \ee
where we have decomposed a $p-$form RR flux $G^{(p)}$ according to
how many legs it does have along the directions of the isometry
group $G$. Since $G$ acts by isometries, the fluxes $F^{(p-a)}, \
a=0,1,\cdots, d$ will have no dependence on the isometry
coordinates $\theta^i$. We map this differential form to a
Clifford algebra element in the usual way: \be \label{cliffordF} F
= \sum_p G^{(p)} = \sum_p \left(F^{(p)} + F_I^{(p-1)} \psi^I +
\frac{1}{2} F_{IJ}^{(p-2)} \psi^I . \psi^J + \cdots + F^{(p-d)}
\psi^1 . \cdots . \psi^d \right). \ee
 The difference we have
here is that it is $\sigma^I$ and not $dx^i$ that we identify with
the Clifford algebra element $\psi^I$, for $I= 1,\cdots, d$. On
the other hand, for $a =d+1, \cdots, 10$, $dx^{a}$ is replaced
with $\psi^{a}$, as usual. Here, $\psi^{\alpha}, \ \alpha = (I,
a)$ are the Clifford algebra elements $\psi^{\alpha} = 1/\sqrt{2}
\Gamma^{\alpha}$, where $\Gamma^{\alpha}$ are the Gamma matrices.
For more details, see \cite{Catal-Ozer:2019hxw}. For index
conventions, see Appendix (\ref{conventions}).

\section{Embedding the homogeneous YB model in DFT}\label{YBinDFT}

In the previous section, we saw that the fields in the target
space of the YB deformed GS sigma model can be obtained by
applying a coordinate dependent $O(d,d)$ matrix on the initial
target space fields. In this section we  describe this $O(d,d)$
transformation as a solution generating transformation in Double
Field Theory (DFT). For this purpose, we start with a quick review
of DFT.

\subsection{A brief review of DFT}

DFT is a field theory defined on a doubled space, which implements
the $O(d,d)$ T-duality symmetry of string theory as a  manifest
symmetry. In addition to the standard space-time coordinates, the
doubled space also includes dual coordinates, which are associated
with the winding excitations of closed strings on backgrounds with
non-trivial cycles. The space-time and dual coordinates $X^M =
(\tilde{x}_{\mu},  x^{\mu}) $ transform as a vector under the
T-duality group $O(d,d)$.

In DFT, the semi-Riemannian metric and the B-field are encoded in
the generalized metric $\cH$, which is an element in $SO^-(d,d)$.
In order to write down the Lagrangian for the RR sector, one needs
the spinor field $\ciftS \in Spin^-(d,d)$ that projects onto $\cH$
under the double covering homomorphism between $Spin(d,d)$ and
$SO(d,d)$, that is $\rho(\ciftS) = \cH$. To describe the p-form
fields in the RR sector, one starts with the democratic
formulation of Type II supergravity. In this formulation, the
p-form fields are packaged in a polyform as in (\ref{spinorF}),
which is then mapped to a spinor field $\chi$ as in
(\ref{cliffordF}). In DFT, this spinor field is allowed to have
dependence both on the space-time coordinates and the winding type
dual coordinates. Another dynamical  field in DFT is the
generalized dilaton field $d$, which  is a generalized scalar
field. A generalized scalar field and more generally generalized
tensor fields are defined according to how they transform under
generalized diffeomorphisms of DFT. Infinitesimal transformations
under generalized diffeomorphisms are generated by the generalized
Lie derivative, which defines the D-bracket that generalizes the
Lie bracket. Anti-symmetrization of the D-bracket gives the
C-bracket. The gauge algebra of infinitesimal transformations
under generalized diffeomorphisms closes under the C-bracket. For
more details see \cite{Catal-Ozer:2017cqr}, \cite{dftRR}.

In its current formulation, DFT is a consistent field theory only
when a certain constraint, called the strong constraint is
satisfied. When the strong constraint is satisfied generalized
tensor fields become sections of the direct sum of the tangent and
the cotangent bundle\footnote{Here, the tangent bundle is the
union of all `physical tangent spaces' determined at each point of
the doubled manifold  by the strong constraint as a maximally
isotropic subspace  of the doubled tangent space. Although the
structure of a doubled manifold and its tangent space is not very
well understood, this choice of a physical tangent space can at
least be done in a local chart.}, generalized diffeomorphisms
reduce to the semi-direct product of space-time diffeomorphisms
and B-field gauge transformations, and the C-bracket becomes the
Courant bracket. This is the framework of generalized geometry
program of Nigel Hitchin
\cite{Hitchin:2004ut},\cite{Gualtieri:2003dx}.  A trivial solution
of the strong constraint occurs when all the fields and gauge
parameters in the theory are independent of  the winding type
coordinates. Such fields are said to belong to the supergravity
frame and it can be shown that the DFT action reduces to the
standard supergravity action in the supergravity frame. Another
well known solution of the strong constraint is provided when  all
the fields except for the dilaton field depend on all but one of
the space-time coordinates and have no dependence on the winding
type coordinates, whereas the dilaton field does also have a
linear dependence on the winding type coordinate dual to the
remaining space-time coordinate. In such a frame, DFT equations
are known to reduce to generalized supergravity equations of
\cite{gse1},\cite{gse2}\footnote{Generalized supergravity
equations (GSE) is a deformation of supergravity equations
determined by a Killing vector field. In an adopted coordinate
system this Killing vector field generates translations along the
space-time coordinate on which the DFT fields have no
dependence.}. Henceforth, we will call this frame the generalized
supergravity frame.

\subsection{YB matrix as a twist matrix in GDFT}\label{YBinGDFT}

From the way they are constructed the target space fields of the
YB deformed GS sigma model form a solution of supergravity
equations, if the corresponding R-matrix is unimodular and of GSE
if not unimodular \cite{Borsato:2018idb}. Then, according to the
discussion above, one can construct DFT fields corresponding to
these target space fields which solve the field equations for DFT
either in the supergravity frame or in the generalized
supergravity frame, depending on the unimodularity of the
R-matrix. On the other hand, the initial target space fields
before the deformation also form a solution for the field
equations of DFT in the supergravity frame, as they belong to the
target space of a GS string sigma model. We showed in
\ref{ybmatrixsection} that the homogeneous YB model is obtained
through the action of the $O(d,d)$ matrix $T_{{\rm YB}}$. Hence,
the YB deformation of the GS sigma model is indeed a solution
generating $O(d,d)$ transformation for the DFT fields
corresponding to the target space fields. A natural question to
ask at this point is as to how  the condition of unimodularity and
the classical Yang-Baxter equation arises on the DFT side. For
addressing this question, we find it useful to utilize the methods
developed in \cite{Catal-Ozer:2019hxw}, where we considered DFT
fields whose dependence on the doubled DFT coordinates are
separated as $\phi(x, Y) = U(Y) . \phi(x)$. Here, the $O(d,d)$
matrix $U(Y)$ is called the twist matrix, $\phi(x, Y)$ denotes a
generic DFT field, $(x, Y)$ are DFT coordinates and the action of
$U(Y) \in O(d,d)$ is determined by how $\phi$ transforms under the
$O(d,d)$ duality group of DFT (or more generally under $Spin(d,d)$
if $\phi$ is a spinor field in DFT). Using terminology from
Scherk-Schwarz compactifications, we call the fields $\phi(x, Y)$
twisted fields, whereas the fields $\phi(x)$ are called  untwisted
fields. We showed in \cite{Catal-Ozer:2019hxw} that  the twisted
fields satisfy the field equations of DFT, if and only if the
untwisted fields satisfy the field equations of Gauged Double
Field Theory (GDFT), provided that  $U(Y)$ satisfies a certain set
of conditions.\footnote{GDFT is a deformation of DFT, obtained
from a Scherk-Schwarz reduction and the deformation is determined
entirely by the so called fluxes associated with the twist matrix
$U(Y)$. For more details, see Appendix \ref{fluxes}.} This
discussion is relevant for analyzing the homogeneous YB model, as
the fields here are also twisted fields with the twist matrix
being $T_{{\rm YB}}$, as was shown in section
\ref{ybmatrixsection}. This perspective will allow us to read off
the CYBE and the condition of unimodularity in terms of fluxes on
the GDFT side.

From the rules (\ref{transformH}) and (\ref{transformF}) we
presented in the  section \ref{ybmatrixsection} and section
\ref{RRfields}, we know that the fields in the target space of the
homogeneous YB model are of the following form\footnote{For now,
we assume that the R-matrix is unimodular. As will be discused in
section \ref{dilaton}, if the R-matrix is not unimodular, the
generalized dilaton field is forced to have a linear dependence on
the winding type coordinates.}:\begin{eqnarray} \label{NATDH}
\cH'^{AB}(x, \nu) &=& (T_{{\rm YB}})^A_{\ C}(\nu)
\cH^{CD}(x) (T_{{\rm YB}})^B_{\ D}(\nu),\\
\label{NATDS} \cK'(x, \nu) &=& S_{{\rm YB}}(\nu)
\cK(x) (S_{{\rm YB}})^{\dagger}(\nu),\\
\label{NATDF} F'(x, \nu) &=& e^{-B'(x, \nu)} S_{{\rm YB}}(\nu)
e^{B(x)} F(x),\\ d'(x, \nu) &=& d(x),
\end{eqnarray} where the fields on the right hand side without prime are the DFT fields constructed
out of the target space fields  before the deformation. The
coordinates $\nu = \{\nu^I\}, I=1, \cdots, d$ are regarded as
coordinates on the group manifold $G$ by using the relation
(\ref{parameters}) derived in  \cite{Borsato:2018idb}. The fields
before and after the deformation have no dependence on the winding
type coordinates, that is, they belong to the supergravity frame.
Also, the fields both before and after the deformation have flat
indices $^A = (_\alpha, \ ^{\alpha})$ with $\alpha = (I, a)$. For
 details on the index conventions, see Appendix
(\ref{conventions}). In (\ref{NATDF}), $\rho(S_{{\rm YB}}) =
T_{{\rm YB}}$ and $B'(x, \nu) $ is read off from the antisymmetric
part of $\cH'(x, \nu)$ in (\ref{NATDH}). The spinor field $\cK$ in
(\ref{NATDS}) is defined as $\cK = C^{-1} \ciftS$ and its
transformation rule is determined by that of $\cH$ since
$\rho(\ciftS) = \cH$. For more details, see
\cite{Catal-Ozer:2017cqr}. Note that the fields $\cH^{AB}(x),
\cK(x), d(x), F(x)$ satisfy the field equations of GDFT with
geometric fluxes $ f_{IJ}^{\ \ K} = C_{IJ}^{\ \ K}$ associated
with the isometry, as discussed in Appendix (\ref{conventions}).

Hence, the DFT fields constructed from the fields in the target
space of the homogeneous YB model are indeed twisted fields, where
the twist matrix is the YB matrix $T_{{\rm YB}}$ in
(\ref{YBmatrix}). The twist matrix $T_{{\rm YB}}$ does  satisfy
the constraints (\ref{cond2}, \ref{cond1}) presented in Appendix
(\ref{fluxes}). Firstly, it is obtained by embedding an $O(d,d)$
matrix in $O(10,10)$ and the untwisted fields do not depend on the
corresponding $d$ coordinates. This ensures that (\ref{cond1}) is
satisfied. Also, $T_{{\rm YB}}$ does not depend on the winding
type coordinates, which then implies that (\ref{cond2}) is also
satisfied. Another condition that needs to be imposed on the twist
matrices for a consistent reduction is that the associated fluxes
defined in (\ref{structure}) should be constant. In subsection
(\ref{fluxcomp}) we will compute these fluxes and show that they
are indeed constant. Moreover, we will show that the CYBE and the
unimodularity condition for the R-matrix become conditions on the
fluxes. As a preparation for this computation, we give below a
brief review of R-matrices and the related algebraic structures.

\subsubsection{Properties of the R-matrix and the dressed
R-matrix}\label{rmatrix}

Let $G$ be a semisimple Lie group with Lie algebra $\mathbf{g}$.
Let $R$ be an endomorphism on $\mathbf{g}$ with $R^{I}_{\ J}$
being the components of the corresponding matrix with respect to a
fixed basis $T_I$ of $\mathbf{g}$:
\begin{equation} RX=(RX)^{I}T_{I}=R^{I}_{\ J} X^{J} T_{I}.
\end{equation} Let $R^{IJ} = \kappa^{IL} R^J_{\ L}$, where $\kappa$ is the
non-degenerate Cartan-Killing form on $\mathbf{g}$. We  assume
that $R$ is skew-symmetric with respect to $\kappa$ so that
$R^{IJ} = - R^{JI}$. The CYBE for the operator R is the following:
\be \label{YB} [RX, RY] - R([RX, Y] + [X, RY]) = 0, \ \ \ \forall
X,Y \in \mathbf{g}. \ee  With respect to the fixed basis $T_I$
this is equivalent to the following equation
\begin{equation} \label{yang2}
R^{LI} R^{MJ} C_{LM}^{\ \ \ K} + R^{LJ} R^{MK} C_{LM}^{\ \ \ I} +
R^{LK} R^{MI} C_{LM}^{\ \ \ J} = 0,
\end{equation} where $C_{IJ}^{\ \ K}$ are the structure constants of the
Lie algebra $\mathbf{g}$ with respect to the basis $T_I$ and all
indices are raised and lowered by the Killing form $\kappa$. The
equation (\ref{YB}) is a sufficient condition for the bracket $[ ,
]_R$ to satisfy the Jacobi identity (see, for example, the book
\cite{book}), where
\begin{equation}\label{Rbracket}
[X,Y]_{R}= [RX, Y]+[X, RY].
\end{equation} Then, this yields a new Lie algebra $\mathbf{g}_R$ with the same
underlying vector space as that of $\mathbf{g}$ and the new
bracket $[ , ]_R$.   The structure constants $\tilde{C}_{IJ}^{\ \
K}$ of $\mathbf{g}_R$ with respect to the basis $T_I$    can be
computed to be
\begin{equation} \label{f tilde}
\tilde{C}_{IJ}^{\ \ L}=C_{KJ}^{\ \ \ L} R^K_{\ I} -C_{KI}^{\ \ \
L} R^{K}_{\ J} =-\tilde{C}_{JI}^{\ \ \ L}.
\end{equation} Lowering the index $L$ and using the fact that $C_{IJK}$ are totally anti-symmetric in all of its indices we obtain
\be \tilde{C}_{IJL} = -C_{KLJ} R^K_{\ I} + C_{KLI} R^{K}_{\ J}.
\ee Now we raise the indices $I$ and $J$ to obtain \be
\label{tildeC} \tilde{C}^{IJ}_{\ \ \ L} = C_{KL}^{\ \ \ J} R^{IK}
- C_{KL}^{\ \ \ I} R^{JK}. \ee Note that \be \tilde{C}^{IJ}_{\ \ \
K} \tilde{T}^K = [\tilde{T}^I, \tilde{T}^J]_{R}, \ee where \be
\tilde{T}^I = \kappa^{IJ} T_J \ee are elements of the dual Lie
algebra $\mathbf{g}_R^*$, where we have identified $\mathbf{g}_R$
and its dual $\mathbf{g}_R^*$ by the bilinear from $\kappa$ in the
usual way.

The dressed matrix is defined as in (\ref{dressed}), which we
rewrite here for convenience: \be R_g = Ad_{g^{-1}} R Ad_g. \ee An
important property that we will use in computing the fluxes is
that the dressed R-matrix is also an R-matrix that satisfies the
CYBE (whenever $R$ does) and as such it defines a   Lie bracket $[
, ]_{R_g}$, which is in fact equivalent to $[ ,
]_R$\footnote{This, of course, is natural as $R_g$ is constructed
by taking an R-matrix $R$ at the identity element of the group
manifold $G$ and by extending it to the whole manifold by the
adjoint action, which is an automorphism of the Lie bracket.}.
Then the equations (\ref{yang2}) and (\ref{f tilde}) are still
true, when one replaces $R$ with $R_g$. These facts can be shown
easily by using the fact that $Ad_g$ is an automorphism of the Lie
bracket.

We showed in (\ref{dressed2}) that the following holds for the
dressed R-operator \be R_g^{IJ} = K^I_{L} R^{LM} K^J_M, \ee where
$K^I_J$ is as in  (\ref{Kdefn}). The coordinate dependence of the
dressed R-matrix comes from the functions $K^I_J$. So, in
computation of the fluxes, we will encounter terms of the form
$\del_L K^I_J$, where $\del_L = l_L^{\ i}
\del_{i},$\footnote{\label{altnot} The reason why we have
derivatives $\del_I$ rather that $\del_i$ becomes clear in the
computation (\ref{formul}).} where $ l^I_{\ i} l_L^{\ i}=
\delta^I_L.$ In order to calculate this, first note that the
functions $K^I_J$ are just \be K^J_I = i_{k_I} \sigma^J, \ee where
$k_I$ are the right-invariant vector fields in (\ref{rightinvvf})
and $\sigma^I$ are the left-invariant one-forms. Now, using the
fact \be \cL_{k_I} \sigma^J = di_{k_I} \sigma^J + i_{k_I}
d\sigma^J  = di_{k_I} \sigma^J + i_{k_I} (\frac{1}{2} C^J_{KL}
\sigma^K \wedge \sigma^L) = 0, \ee we immediately obtain \be
\label{Kturev} \del_L K^I_J = - C_{PL}^{\ \ \ I} K^P_J. \ee

\subsubsection{Fluxes associated with the YB matrix and the
CYBE}\label{fluxcomp}

We are now ready to compute the fluxes associated with the twist
matrix (\ref{YBmatrix}). We have \be (T_{{\rm YB}})^A_{\ B} =
\left(\begin{array}{cc}
                    T_{\alpha}^{\ \beta} & T_{\alpha \beta} \\
                    T^{\alpha \beta} & T^{\alpha}_{\ \beta}
                    \end{array}\right), \ee with $T_a^{\ b} = \delta_a^{\
                    b}, \ T^a_{\ b} = \delta^a_{\ b}, \ T_a^{\ J}
                    = T^a_{\ J} = T_I^{\ b} = T^I_{\ b} = 0$. For more details on the index structure, see Appendix \ref{conventions}. The
                    embedded $d \times d$ matrix, which we have
                    also called the YB matrix  has components
                    $T^I_{\ J} = \delta^I_{\ J}, \ T_I^{\ J} =
                    \delta_I^{\ J}, \ T_{IJ} = 0$ and $T^{IJ} =
                     \Theta^{IJ} = \eta R_g^{IJ}$. Then, the formula  in (\ref{structure}) for the fluxes
                    becomes \be f_{ABC} = 3
\Omega_{[ABC]}, \ \ \ \Omega_{ABC} = -T^E_{\ A}
                    \del_E
T^F_{\ B} (T^{-1})^D_{\ F}\eta_{CD}. \ee We expand this as \bea
\nonumber \Omega_{ABC} &=& -T^E_{\ A}
                    \del_E
T^F_{\ B} (T^{-1})^D_{\ F}\eta_{CD}  \\
\nonumber &=& -T^{\alpha}_{\ A}
                    \del_{\alpha}
T^F_{\ B} (T^{-1})^D_{\ F}\eta_{CD}  \\ &=&-T^{I}_{\ A}
                    \del_{I}
T^J_{\ B} (T^{-1})^D_{\ J}\eta_{CD}. \label{formul} \eea In the
second line we used the fact that the twist matrix $T_{{\rm YB}}$
has no dependence on the winding type coordinates and in the third
line we used the fact that the twist matrix depends only on the
coordinates along the isometry directions. (Recall that the
doubled coordinate $A$ decompose as $^A = (_{\alpha}, \ ^\alpha)$,
where $\alpha = (I, a)$ with the indices $I$ labelling the
isometry directions.)

 Using the properties of the $O(d,d)$ matrices in general (see Appendix \ref{app2}), and the structure of the YB matrix in particular, we  find that the only
non-vanishing fluxes are \be R^{IJK} = 3\Omega^{[IJK]} = 3
\Theta^{L[I} \del_L \Theta^{JK]}, \ee and \be \label{Qflux1}
 Q_I^{\ JK} = \Omega_I^{\ JK} + \Omega^{JK}_{\ \ I} + \Omega^{K \
J}_{\ I} = \Omega_I^{\ JK} =  \del_I \Theta^{JK}. \ee Note that
the second equality in (\ref{Qflux1}) above holds because the
components $ \Omega^{JK}_{\ \ I}$ and $ \Omega^{K \ J}_{\ I} $ are
zero  due to the structure of the YB matrix (\ref{YBmatrix}). Now,
using equations (\ref{Kturev}) and (\ref{dressed2}) we find \be
\Omega^{IJK} = \eta^2 R_g^{LI} R_g^{RJ} C_{RL}^{\ \ K} - \eta^2
R_g^{LI} R_g^{RK} C_{RL}^{\ \ J}, \ee which gives \be
\label{Rflux} R^{IJK} = \Omega^{IJK} + \Omega^{JKI} + \Omega^{KIJ}
= \eta^2 R_g^{L[I} R_g^{\mid R \mid J} C_{RL}^{\ \ K]}. \ee We can
also compute the Q-flux by using (\ref{Kturev}) and
(\ref{dressed2}): \be \label{Qflux} Q_I^{\ JK} = \Omega_I^{\ JK} =
\eta C_{PI}^{\ \ J} R_g^{KP} - \eta C_{PI}^{\ \ K} R_g^{JP} = \eta
C_{PI}^{\ \ [J} R_g^{K]P}. \ee

Now recall that $R^{IJ}$ are the components of an R-matrix that
satisfies the CYBE (\ref{yang2}) and the relation (\ref{tildeC}).
As a result, the dressed R-matrix $R_g$ (\ref{dressed}) also
satisfies (\ref{yang2}) and (\ref{tildeC}). Using these in
(\ref{Rflux}) and (\ref{Qflux}) respectively we obtain \bea
\label{finalflux1} R^{IJK} &=& 0, \\ \label{finalflux2}  Q_I^{\
JK} &=& \eta \tilde{C}^{JK}_{\ \ I}. \eea

Therefore, we conclude that YB deformation is a process which
takes a solution  of DFT with geometric flux associated with the
isometry group $G$ and deforms it to another solution of DFT with
vanishing R-flux and non-vanishing Q-flux. The non-vanishing
Q-flux is given by the structure constants of the dual Lie algebra
$\mathbf{g}_R^*$ of $\mathbf{g}_R$ determined by the R-matrix $R$.
This is in agreement with the results of
\cite{Fernandez-Melgarejo:2017oyu}, which finds that the YB
deformation of the Minkowski space and the background $AdS_5\times
S^5$  induces  Q-flux on the resulting backgrounds, arising from a
non-trivial monodromy in the so called $\beta$-field around closed
cycles. What we find here by using the framework of GDFT agrees
with their result and extends it for YB deformation of any GS
sigma model. Also, using the framework of GDFT reveals the role of
the Q-flux as the structure constants of a Lie algebra dual to the
isometry algebra. The Q-flux of deformed backgrounds and its role
as
 the dual structure constants to geometric flux is
 also studied in \cite{Lust:2018jsx}, but in the context of Poisson Lie
 duality of sigma models. However, for the YB deformation, this
 feature of the deformed models cannot be captured in
 \cite{Lust:2018jsx}.

Before we move on to the discussion of the non-unimodularity
condition for the R-matrix in the next subsection, we would like
to make a comment. In obtaining the equation (\ref{finalflux1}),
we used the fact that if the CYBE (\ref{yang2}) is satisfied by
the  R-matrix (and hence by the dressed R-matrix), then the R-flux
computed in (\ref{Rflux}) vanishes. In fact, the converse
statement is also true: if we demand the R-flux to
vanish\footnote{Note that this is in fact a reasonable thing to
demand, as the non-vanishing R-flux is usually a sign of
non-geometry, even locally.}, then the CYBE must be satisfied by
the R-matrix that determines the YB deformation. Indeed, using the
fact that the structure constants $C_{IJ}^{\ \ K}$ are
antisymmetric in their lower indices we see
 that the equation $ R^{IJK} = 0 $ for the R-flux is
equivalent to the CYBE (\ref{yang2}) for the dressed R-matrix.


\subsection{The Dilaton Field and the Generalized Supergravity
Equations}\label{dilaton}

In this section, we analyze the case when the R-matrix that
determines the YB deformation is non-unimodular. As we discussed
in the introduction, unimodularity of the R-matrix is required for
the homogeneous YB model to be a solution of supergravity
equations. On the other hand, when it is non-unimodular, the model
is known to be a solution of GSE \cite{Borsato:2016ose}.

The condition for unimodularity presented in
\cite{Borsato:2016ose} is\footnote{This also means that the
quasi-Fobenius subalgebra of $\mathbf{g}$ determined by $R$ is
unimodular, see \cite{Borsato:2016ose}.}: \be
\label{unimodularity} R^{IJ} C_{IJ}^{\ \ K} = 0.\ee Since the Lie
algebra $\mathbf{g}$ we start with is unimodular so that
$C_{IJ}^{\ \ I} = 0$ in any basis, a careful examination of
(\ref{tildeC}) shows that the condition (\ref{unimodularity}) is
satisfied if and only if the structure constants
$\tilde{C}^{IJ}_{\ \ K}$ are traceless. This then implies that the
R-matrix is unimodular if and only if the Q-flux generated by the
YB matrix is traceless, see (\ref{finalflux2}). This is in
agreement with the results of \cite{Fernandez-Melgarejo:2017oyu},
which also finds that the non-unimodularity of the R-matrix is
measured by the trace of the Q-flux.

Now, assume that  condition (\ref{unimodularity}) is not met, that
is, the R-matrix is not unimodular. As a result, the Q-flux is
trace-full. Looking at (\ref{fa}), one immediately sees that the
trace of the Q-flux is just the fluxes $f^I$ defined there.
Indeed, \be \label{fa2} f^I = \Omega^{AI}_{\ \ A} = \Omega^{\alpha
I}_{\ \ \alpha} + \Omega_{\alpha}^{\ I \alpha} = \Omega_{J}^{\ IJ}
= Q_{J}^{\ IJ}, \ee since all other components of $\Omega$ are
 zero due to the structure of the YB matrix. Note that this means, by using (\ref{finalflux2}) that \be \label{fa3} f^I = \eta \tilde{C}^{IJ}_{\ \ J}.\ee
  As a result, when the
R-matrix is not unimodular, the fluxes $f^I$ are non-vanishing and
this contributes to $\eta^I$, whose definition is given in
(\ref{structure}). However, it is well-known that the GDFT action
with non-vanishing $\eta_A$ is not consistent. This follows from
the requirement of gauge invariance of  the GDFT action. In the
case of the NS-NS sector, this analysis was made in
\cite{Grana:2012rr}. For the RR sector, it was shown in
\cite{Catal-Ozer:2017cqr} that the RR sector of the GDFT action is
gauge invariant, only when the fluxes $\eta^A$ vanish, see
equation (4.66) there. In order to satisfy this condition,
 the $f^I$ part in (\ref{structure}) should be
compensated by a non-trivial dilaton anzats. A similar situation
was considered in \cite{Catal-Ozer:2019hxw} and
\cite{Catal-Ozer:2017ycb}. Rewriting (\ref{structure}) in
components, we see that we need to have \bea \label{gse1} \eta^I &
= & -f^I - 2(U^{-1})^{MI}
\del_M \sigma = 0, \\
\eta_I & = & -f_I - 2(U^{-1})^M_{\ \ I} \del_M \sigma = 0. \eea
When the twist matrix is equal to the YB matrix $U^{-1} = T_{{\rm
YB}}$, we have $(U^{-1})^{IJ} = \eta \Theta^{IJ}$, and
$(U^{-1})^I_{\ J} = \delta^I_{\ J}, \ (U^{-1})_I^{\ J} =
\delta_I^{\ J}, \ (U^{-1})_{IJ} = 0$. Plugging this in
(\ref{gse1}) we get \bea \label{gse3} -2 \eta \Theta^{JI} \del_J
\sigma - 2 \delta_J^{\ \ I} \tilde{\del}^J \sigma &=& f^I = {\rm
constant}, \\
\label{gse2} -2\delta^J_{\ \ I} \del_J \sigma &=& 0. \eea
Combining (\ref{gse2}) and (\ref{gse3}) we obtain
$$ \del_I \sigma = 0, \ \ \ \tilde{\del}^I \sigma = -f^I / 2 = {\rm
constant}.$$ In other words, $\sigma$ is linear in the  winding
type  coordinates and does not depend on the standard coordinates.
Then, the generalized dilaton field is of the form: \be
\label{linearanzats} d(x, \tilde{\nu}) = d(x) - \frac{f^I}{2} \
\tilde{\nu}_I. \ee

Due to the linear dependence of the generalized dilaton field on
the winding type coordinates\footnote{Note that, due to the
appearance of the term $e^{-\sigma(Y)}$ in the anzats
(\ref{anzatschi}), the spinor field $F = e^{-B} \slashed{\del}
\chi$ also has a dependence on the winding type coordinates. The
other DFT fields $\cH$ and $\ciftS$ have no dependence on winding
type coordinates. For more details on a similar situation, see
\cite{Catal-Ozer:2017ycb}.} we are now in the generalized
supergravity frame discussed at the end of section \ref{YBinDFT}.
It was shown in \cite{Sakatani:2016fvh} and
\cite{Sakamoto:2017wor} that  the field equations of DFT in this
frame reduce to GSE, where  the components $I^m$ of the Killing
vector field $I$ that defines the deformation in GSE is identified
with $d^J$, where \be \label{mdft1} d(x, \tilde{\nu}) = d(x) + d^J
\tilde{\nu}_J, \ \ \ {\rm and} \ \ \ I^J = d^J,\ee see equation
(3.51) of \cite{Sakamoto:2017wor}. When we compare
(\ref{linearanzats}) with (\ref{mdft1}), we see that the
components of the Killing vector field $I$ are given by $-f^I / 2$
in our formalism. Note that this is in agreement with the results
of \cite{Demulder:2018lmj} (see equation (3.63) in that paper),
but disagrees with a factor of 2 with the result $I^i =
f^{j}_{ij}$ of \cite{Hong:2018tlp} obtained for NATD. The source
of this discrepancy was discussed and resolved in
\cite{Sakatani:2019jgu}, as will discuss shortly.

Let us also discuss  the consistency of our results with those of
\cite{Araujo:2017enj} and \cite{Araujo:2017jkb}. Looking at
(\ref{Qflux1}), one sees that the trace of the Q-flux is
equivalent to the divergence of $ \Theta$: \be \label{yenidenklem}
Q_K^{\ IK} =  \del_K ( \Theta^{IK}), \ee where $\Theta^{IK} = \eta
(R_g)^{IK}$ is given in (\ref{dressed2}). Recall we discussed in
section \ref{ybmatrixsection} that $ \Theta^{IK}$ is in fact
equivalent to the noncommutativity parameter $\Theta_{NC}$ of the
papers \cite{Bakhmatov:2017joy}, \cite{Bakhmatov:2018apn} (see
footnote (\ref{onemlidipnot2})). In the papers
\cite{Araujo:2017enj}, \cite{Araujo:2017jkb} the divergence of
$\Theta_{NC}$ is identified with the Killing vector field $I$ that
defines the GSE, see equation (3) in \cite{Araujo:2017jkb}. In our
approach based on the framework of GDFT, (\ref{yenidenklem})
equates the divergence of $\Theta^{IK}$ to the trace of the
Q-flux, which arises as a measure of the non-unimodularity of the
R-matrix. On the other hand, as we discussed above, the trace of
the Q-flux is equivalent to $f^I$, which are the type of fluxes
that force the YB deformed fields to lie in the generalized
supergravity frame in which the field equations of DFT reduce to
GSE with $ I^I = -\frac{1}{2} f^I$. To summarize, we have the
following relations in our framework: \be \label{birdenklemdaha}
I^I = - \frac{1}{2} f^I = -\frac{1}{2} Q_K^{\ IK} = - \frac{1}{2}
\del_K (\Theta^{IK}) =\frac{1}{2} \del_K ( \Theta^{KI}). \ee
Comparing (\ref{birdenklemdaha}) with $I^I = \nabla_K
\Theta_{NC}^{KI}$ presented as equation (3) of
\cite{Araujo:2017jkb}, we see that
 there is a mismatch with a factor of 2.   This discrepancy is in fact caused by the
factor 2 mismatch  mentioned at the end of the previous paragraph
between the results for the components of the Killing vector field
$I$ found in \cite{Demulder:2018lmj} (and here) and in
\cite{Hong:2018tlp}. A resolution for this discrepancy  is
introduced in \cite{Sakatani:2019jgu}\footnote{We  thank the
anonymous referee for informing us on the discussion  in
\cite{Sakatani:2019jgu} on how to resolve the discrepancy between
the results of \cite{Demulder:2018lmj} and of
\cite{Hong:2018tlp}.}, which involves a field redefinition for the
vector field $I$, as described in equation (6.67) of
\cite{Sakatani:2019jgu}. They redefine the vector field $I$ so
that  $I^J = d^J \rightarrow I^J = \frac{1}{2} \eta
\tilde{C}^{IJ}_{\ \ I} + d^J$ , where $d^J$ is as in
(\ref{mdft1})\footnote{We have rewritten equation (6.67) of
\cite{Sakatani:2019jgu} in our conventions. We expand the Killing
vector field $I$ with respect to left invariant vector fields so
that $I = I^J \del_J = I^J l^i_{\ J} \del_i$. What we call $l^i_{\
J}$ here is called $v_a^m$ in \cite{Sakatani:2019jgu}. Also, we
need the deformation parameter $\eta$ in our transformation,
whereas it is not needed in \cite{Sakatani:2019jgu}, where they
work in the setting of Poisson Lie T-duality.}. Then, $d^J = I^J -
\frac{1}{2} \eta \tilde{C}^{IJ}_{\ \ I} = I^J + \frac{1}{2} f^J,$
where we have used (\ref{fa3}). Comparing it with equation
(\ref{linearanzats}) above, we see that this gives $ -\frac{1}{2}
f^J = d^J = I^J + \frac{1}{2} f^J$ which gives $I^J = -f^J$ as in
equation (6.71) of \cite{Sakatani:2019jgu}. Plugging this in
(\ref{birdenklemdaha}), we obtain $$I^I = \del_K \Theta^{KI},
$$ which is exactly the same  as equation (3) of
\cite{Araujo:2017jkb}\footnote{Recall that $\del_K = l^i_{\ K}
\del_i$, which is indeed a covariant derivative on a group
manifold.}.

Recall that the fields in the homogeneous YB model are known to
form a solution of GSE, when the R-matrix is not unimodular. The
discussion above shows that for this case, the deformed  fields
are forced to lie in the frame in which these equations can be
embedded in field equations of DFT.  Hence, viewing  YB
deformation as a solution generating transformation in DFT
provides a nice framework in which the unimodular and
non-unimodular cases can be considered simultaneously.

\section{Conclusion}\label{conclusion}

In this paper, we studied the homogeneous YB deformation proposed
in \cite{Borsato:2018idb} for a generic GS sigma model. We showed
that the deformation in the target space fields, including the RR
fields is generated by a non-constant $O(d,d)$ transformation,
generalizing previous results obtained by \cite{Sakamoto:2017cpu}
and  \cite{Sakamoto:2018krs}. The associated matrix, which we call
the YB matrix, is determined by an R-matrix satisfying the CYBE.
The coordinate dependence comes from extending this fixed R-matrix
to a dressed R-matrix on the whole group manifold (of the isometry
group) by the adjoint action. Since the adjoint action is an
automorphism of the Lie bracket, the resulting dressed R-matrix
also satisfies the CYBE, whenever the initial R-matrix does.
Relating this to the open-closed string map approach of
\cite{Araujo:2017jkb},\cite{Araujo:2017jap},\cite{Araujo:2017enj},
we see that their bi-Killing anzats for the noncommutativity
parameter arises  naturally from this construction with the
adjoint action.

The YB matrix  is related by a constant $O(d,d)$ transformation to
the matrix found in \cite{Catal-Ozer:2019hxw} that generates NATD
backgrounds, solidifying the relation between NATD and YB
deformations. It also provides a nice framework to see how the
homogeneous  YB deformation generalizes the Lunin-Maldacena
deformation, generated by TsT transformations. Note that the TsT
transformation was generalized in \cite{CatalOzer:2009xd} to a
solution generating transformation for 11 dimensional
supergravity. Using the formalism there and the results we obtain
here, we expect it to be possible to also generalize the YB
deformation to a solution generating transformation in 11
dimensions.  We would like to note that generalized YB
deformations in 11 dimensions has been studied very recently
\cite{eoin} within the framework of Exceptional Field Theory, a
formalism which provides a U-duality covariant formulation for 11
dimensional supergravity.

Besides making calculations  easier, our approach gives a natural
embedding of the homogeneous YB model  in DFT. This enabled us to
show that the YB deformed fields can be regarded as duality
twisted fields in the context of GDFT. We computed the fluxes
associated with the YB matrix and showed that the conditions on
the R-matrix determining the YB deformation manifested themselves
as conditions on the fluxes on the GDFT side. More precisely, we
showed that YB deformation is a process which takes a solution of
DFT with geometric flux associated with the isometry group $G$ and
deforms it to another solution of DFT with vanishing R-flux and
non-vanishing Q-flux given by the structure constants of the dual
Lie algebra determined by the R-matrix.  We also showed that for
YB deformations based on a  non-unimodular R-matrix, the deformed
fields are forced to lie in a frame in which the generalized
supergravity equations can be embedded in  DFT. Hence, viewing  YB
deformation as a solution generating transformation in DFT
provides a nice framework in which the unimodular and
non-unimodular cases can be considered simultaneously. Our
approach based on GDFT also enabled us  to discuss the relation
between the non-unimodularity of the R-matrix, the trace of the
Q-flux and the generalized supergravity equations. We hope that
the approach we pursue here contributes towards a better
understanding of the exciting relation between the CYBE and
(generalized) supergravity.

\section*{Acknowledgments}
We would like to thank E.~O Colgain, N.~S.~Deger and
M.~M.~Sheikh-Jabbari for comments and discussions. This work has
been partially supported by the Turkish Council of Research and
Technology (T\"{U}B\.{I}TAK) through the ARDEB 1001 project with
grant number 114F321. We also acknowledge the financial support of
Scientific Research Coordination Unit of Istanbul Technical
University (ITU BAP) under the project TGA-2018-41742.

\appendix

\section{O(D,D) and Its Action on Curved Backgrounds}\label{app2}

Let $T$ be a matrix in $O(d,d,R)$. Then \be \label{oddefinition} T
\ \eta \ T^t = \eta  \ \ \ {\rm and} \ \ \ T^t \eta T = \eta, \ee
where $\eta$ is the $O(d,d)$ invariant metric with \be \nonumber
\eta =  \left(\begin{array}{cc} 0 & I_d \\
                        I_d & 0 \end{array}\right), \ee where
                        $I_d$ is the $d \times d$ identity matrix.
                        Writing the equations above in indices
                        for the YB matrix and its inverse we get: \be \nonumber (T_{{\rm YB}})^A_{\ C}
\ \eta^{CD} \ (T_{{\rm YB}})^B_{\ D} = \eta^{AB}, \ \ \ \ \
(T_{{\rm YB}}^{-1})^C_{\ A} \ \eta_{CD} \ (T_{{\rm YB}}^{-1})^D_{\
B} = \eta_{AB}. \ee In indices $\eta$ has the following form: \be
\nonumber \eta^{AB} = \left(\begin{array}{cc} \eta_{\alpha \beta} & \eta_{\alpha}^{\ \beta} \\
                        \eta^{\alpha}_{\ \beta} & \eta^{\alpha \beta} \end{array}\right) = \left(\begin{array}{cc} 0 & \delta_{\alpha}^{\ \beta} \\
                        \delta^{\alpha}_{\ \beta} & 0 \end{array}\right). \ee

\noindent If we write   $T$ as  \be \nonumber T =
\left(\begin{array}{cc} a & b \\
                        c & d \end{array}\right) \in O(d,d). \ee  then, (\ref{oddefinition}) implies \be \nonumber a^t c +
                        c^t a = 0, \ \ b^t d + d^t b = 0, \ \ a^t
                        d + c^t b = I. \ee
T can be embedded in $O(D,D,R)$ as follows \cite{Giveon:1994fu}:
\be
 \hat{T} =
\left(\begin{array}{cc} \label{matrixT} \hat{a} & \hat{b} \\
                        \hat{c} & \hat{d} \end{array}\right), \ee
                        where $\hat{a}, \hat{b}, \hat{c}, \hat{d}$
                        are $D \times D$ matrices defined below:

\be \label{embedding} \hat{a} = \left(\begin{array}{cc} a & 0 \\
                                      0 & I \end{array}\right), \
                                      \ \hat{b} = \left(\begin{array}{cc} b & 0 \\
                                      0 & 0 \end{array}\right), \
                                      \ \hat{c} = \left(\begin{array}{cc} c & 0 \\
                                      0 & 0 \end{array}\right), \
                                      \ \hat{d} = \left(\begin{array}{cc} d & 0 \\
                                      0 & I \end{array}\right).
                                      \ee The  action of the $O(D,D)$ matrix $\hat{T}$ on the background
matrix $E$ is defined as below \cite{Giveon:1994fu}: \be
\label{finalE} E'(g', B') = \hat{T} . \ E(g, B) \equiv (\hat{a} E
+ \hat{b})(\hat{c} E + \hat{d})^{-1}. \ee The transformed metric
and the transformed B-field are read off from $E'$ as \be g' =
\frac{E' + E'^t}{2}, \ \ B' = \frac{E' - E'^t}{2}. \ee

\section{GDFT and the Fluxes}\label{fluxes}
GDFT is obtained from a Scherk-Schwarz reduction of DFT with the
following duality twisted reduction anzats:
  \cite{Catal-Ozer:2017cqr}: \bea \label{anzats} \cH^{MN}(X, Y) &=&
(U^{-1})^M_{\ A}(Y) \cH^{AB}(X) (U^{-1})^N_{\ B}(Y), \\
\ciftS(X, Y) &=& (S^{-1})^{\dag}(Y) \ciftS(X) S^{-1}(Y), \\
\label{anzatschi} F(X, Y) &=& e^{-\sigma(Y)} e^{-B(X, Y)} S(Y)
e^{B(X)} F(X), \\
 \label{anzatsdilaton} d(X, Y) &=& d(X) + \sigma(Y). \eea
Here, $X$ denote collectively the coordinates of the reduced
theory. The $Y$  coordinates are the internal coordinates, which
will be integrated out eventually. One can further decompose these
coordinates into dual and standard coordinates as $Y = (\tilde{y},
y)$ and $X = (\tilde{x}, x)$.  The twist matrix $U(Y)$ is an
element of $O(D,D)$ and $\rho(S) = U^{-1}$,  where $\rho$ is the
double covering homomorphism: $\rho:Pin(D,D) \rightarrow O(D,D)$.
$B(X, Y) $ in (\ref{anzatschi}) is read off from the antisymmetric
part of $\cH(X, Y)$ in (\ref{anzats}).

The reduced theory is determined by the so called fluxes $f_{ABC},
\ \eta_A$, which are defined as below: \be \label{structure}
f_{ABC} = 3 \Omega_{[ABC]}, \ \ \ \eta_A =
\partial_M  (U^{-1})^M_{\ A} - 2(U^{-1})^M_{\ A} \partial_M \sigma, \ee  \be \label{omega}
\Omega_{ABC} = -(U^{-1})^M_{\ A} \del_M (U^{-1})^N_{\ B} U^D_{\
N}\eta_{CD}.  \ee  Note that $\Omega_{ABC}$ are antisymmetric in
the last two indices: $\Omega_{ABC} = - \Omega_{ACB}$. We also
make the following definition \be \label{fa} f_A = -\partial_M
(U^{-1})^M_{\ A}  = \Omega^C_{\ AC}. \ee For a consistent
reduction, constraints that should be obeyed by the twist matrices
are as follows \cite{Geissbuhler:2011mx}-\cite{Grana:2012rr}: \bea
\label{cond2}
 \partial^P (U^{-1})^M_{\ A} \partial_P g(X) &=& 0, \\
 \label{cond1}
 (U^{-1})^M_{\ A} \partial_M g(X) &=& \partial_A g(X), \eea
 where $g$ is
 any of the DFT fields ($\cH, \ciftS, \chi$). In addition to the above, the weak and the strong constraint has
to be imposed on the external space so that \be \label{external}
\partial_A \partial^A V(X) = 0, \ \
\partial_A V(X) \partial^A W(X) = 0, \ee for any fields or gauge
parameters $V, W$ that has dependence on the coordinates of the
external space only. Also,   all fluxes must be constant for the
consistency of the reduced theory, which ensures that the $Y$
dependence is completely integrated out in the reduced theory.

\section{Index Structure for the Twist Matrices}\label{conventions}
Our index conventions are as follows:

$M, N, \cdots $: Doubled coordinates; $ ^M = (_{\mu}, \ ^{\mu})$,

$A, B, \cdots $: Doubled coordinates; $ ^A = (_{\alpha}, \
^{\alpha})$,

$\mu = (i, m), \ \ \ \mu = 1, \cdots, 10; \ \ i=1, \cdots, d, \ \
d= {\rm dim} G$,

 $\alpha = (I, a), \ \ \ \alpha = 1, \cdots, 10; \
\ I=1, \cdots, d$.

\noindent  According to the embedding rules in (\ref{embedding}),
a twist matrix $T \in O(D,D,R)$, which only twists the $d$
isometry directions is of the following form: \be T^M_{\ A} =
\left(\begin{array}{cc}
                    T_{\mu}^{\ \alpha} & T_{\mu \alpha} \\
                    T^{\mu \alpha} & T^{\mu}_{\ \alpha}
                    \end{array}\right), \ee with $T_m^{\ a} = \delta_m^{\
                    a}, \ T^m_{\ a} = \delta^m_{\ a}, \ T_m^{\ I}
                    = T^m_{\ I} = T_i^{\ a} = T^i_{\ a} = 0$.

There are  two types of twist matrices relevant for this paper:
$L$ and $T_{{\rm YB}}$. Let us discuss the index structure of
these matrices.

The twist matrix $L$ arises when the Lie group $G$ acts on the
background by isometries. In this case we can write
\begin{eqnarray} ds^2 &=& G_{\mu \nu} dx^{\mu} dx^{\nu} = G_{mn} dx^{m} dx^{n} + 2 G_{mi} dx^{m} d\theta^{i} + G_{ij} d\theta^{i}
d\theta^{j} \label{metric1} \\
& = & G_{mn} dx^{m} dx^{n} + 2 G_{mI} dx^{m} \sigma^I + G_{IJ}
\sigma^I \sigma^J \label{metric2} \\
& = & G_{\alpha \beta} \sigma^{\alpha} \sigma^{\beta},
\label{metric3}
\end{eqnarray} where $\theta^{i}, i= 1, \cdots, d$ are coordinates for $G$ and $\sigma^a = \delta^a_{\ m} dx^m$ and $\sigma^I, I = 1, \cdots, dimG $
are the left invariant 1-forms on  $G$; $\sigma^I = l^I_{\ i}
d\theta^{i}$. Similarly, \bea \label{B1} B &=& \frac{1}{2} B_{\mu
\nu} dx^{\mu} \wedge dx^{\nu} \\
\label{B2} &=& \frac{1}{2} B_{mn} dx^{m} \wedge dx^{n} + B_{mI}
dx^{m} \wedge \sigma^I + \frac{1}{2} B_{IJ} \sigma^I \wedge
\sigma^J. \eea Since the group $G$ acts on the background by
isometries, all the $\theta$ dependence of the fields are encoded
in $l^I_{\ i}$. We define the matrices $G(x, \theta), G(x), B(x,
\theta)$ and $B(x)$ from \be \label{denklem1} ds^2 = d\textbf{x}^T
G(x, \theta) d\textbf{x} = \sigma^T G(x) \sigma, \ \ \ \ B =
d\textbf{x}^T B(x, \theta)\wedge d\textbf{x} = \sigma^T B(x)
\wedge \sigma, \ee where $\wedge$ denote the obvious wedge product
of matrices and $d\textbf{x}$ and $\sigma$ denote the 10-vectors
with components $(dx^1, \cdots, dx^{10})$ and $(\sigma^1, \cdots,
\sigma^d, dx^{d+1}, \cdots, dx^{10})$, respectively. Then the
background matrix $E = G + B$ has the following form: \be E(x,
\theta) = l^T(\theta) E(x) l(\theta), \ee where $l$ is the
$GL(10)$ matrix obtained by embedding in the  $GL(d)$ matrix $l_d$
with components $(l_d)^I_{\ i} = l^I_{\ i}$. The embedding is as
described above so that $(l_d)^I_{\ m} = l^a_{\ i} = 0$ and
$(l_d)^a_{\ m} = \delta^a_{\ m}$. This is equivalent to the
following $O(10,10)$ action : \be \label{separatedE}
 E(x, \theta) =L(\theta). E(x),\ee where $L$ is the following $O(10,10)$
 matrix:
\be \label{geomtwist} L = \left(\begin{array}{cc}
                                  l^T & 0 \\
                                  0 & l^{-1}
                                  \end{array}\right). \ee
One can show  that (\ref{separatedE}) is equivalent to
\cite{ZwiebachL}:\be \label{separatedH}  \cH(x, \theta)= L(\theta)
\cH(x) L^T(\theta). \ee Since the twist matrix $L$ operates
between curved and flat indices, the index structure of it is as
follows: \be \label{separatedH2} \cH^{MN}(x, \theta)=
L(\theta)^M_{\ A} \cH^{AB}(x) L^N_{\ B}(\theta), \ee where we have
identified \be \cH \longleftrightarrow \cH^{MN}
\longleftrightarrow
\left(\begin{array}{cc} G - B G^{-1} B &  G^{-1} \\
                        - G^{-1} B & G^{-1} \end{array}\right).
                        \ee From (\ref{geomtwist}) we read off
                        $L_m^{\ a} = \delta_m^{\ a}, \ L^m_{\ a} =
                        \delta^m_{\ a}, \ L_m^{\ I} =  L^m_{\ I} =
                        L_i^{\ a} = L^i_{\ a} = 0$ and $L_i^{\ I}=
                        (L_d)_i^{\ I} = l_i^{\ I}, \ L^i_{\ I}
                        =(L_d)^i_{\ I} = l^i_{\ I}$, where $l^i_{\
                        I} l^I_{\ j} =  \delta^i_{\ j}$.

One can show that (see \cite{CatalOzer:2005mr}) (\ref{separatedH})
implies the following for the other DFT fields:
 \bea \label{anzatsK} \cK(x, \theta) &=&
S_L(\theta) \cK(x) S_L^{-1}(\theta), \\
\label{separatedF} F(x, \theta) &=& S_L(\theta) F(x) = e^{B(x,
\theta)} S_L(\theta) e^{-B(x)} F(x), \eea where $\rho(S_L) = L$.

From the formulas (\ref{structure}) with $(U^{-1})^M_{\ A} =
L^M_{\ A}$, one finds the the only non-vanishing flux is the
geometric flux \begin{equation}
f_{IJ}^{\;\;\;\;\;K}=-l^{i}_{\;I}\partial_{i}l^{j}_{\;J}l^{K}_{\;j}+l^{i}_{\;J}\partial_{i}l^{j}_{\;I}l^{K}_{\;j}=C_{IJ}^{\;\;\;\;\;K}.
\end{equation} This follows  from the fact that $\sigma^I = l^I_{\ i}
d\theta^i$ are left-invariant one-forms and as such  they satisfy
\be d\sigma^I = \frac{1}{2} C_{IJ}^{\ \ K} \sigma^J \wedge
\sigma^K. \ee

Now, let us discuss the index structure of the twist matrix
$T_{{\rm YB}}$ (\ref{YBmatrix}). First, recall that $\Theta^{IJ} =
l^I_{\mu} \Theta^{\mu \nu} l^J_{\nu}$ where $\Theta^{\mu \nu}
=\eta  k_J^{\mu} R^{JL} k_L^{\nu}$. Then we can write
 \bea T_{{\rm YB}} &=& L^{-1} T_{{\rm YB}}^{\ {\rm curved}} L, \\  \left(\begin{array}{cc} 1 & 0 \\
                                                 \Theta^{IJ} &
                                                1\end{array}\right) &=& \left(\begin{array}{cc}  (l^{-1})^T & 0 \\
                                                0 &
                                                l\end{array}\right) \left(\begin{array}{cc} 1 & 0 \\
                                                \Theta^{\mu \nu} &
                                                1\end{array}\right) \left(\begin{array}{cc} l^T & 0 \\
                                                0 &
                                                l^{-1}\end{array}\right).
                                                \eea
Hence, the index structure is as follows \be (T_{{\rm YB}})^A_{\
B}
 = (L^{-1})^A_{\ M} (T_{{\rm YB}}^{\ {\rm curved}})^M_{\ N} L^N_{\ B}. \ee
Then, \be (T_{{\rm YB}})^A_{\ B} = \left(\begin{array}{cc}
                    T_{\alpha}^{\ \beta} & T_{\alpha \beta} \\
                    T^{\alpha \beta} & T^{\alpha}_{\ \beta}
                    \end{array}\right), \ee with $T_a^{\ b} = \delta_a^{\
                    b}, \ T^a_{\ b} = \delta^a_{\ b}, \ T_a^{\ J}
                    = T^a_{\ J} = T_I^{\ b} = T^I_{\ b} = 0$. The
                    embedded $d \times d$ matrix, which we have
                    also called the YB matrix  has components
                    $T^I_{\ J} = \delta^I_{\ J}, \ T_I^{\ J} =
                    \delta_I^{\ J}, \ T_{IJ} = 0$ and $T^{IJ} =
                     \Theta^{IJ}$.

\end{document}